\documentclass[aps,prl,twocolumn,superscriptaddress,citeautoscript]{revtex4}

\usepackage{graphicx}
\usepackage{amssymb,amsmath}
\usepackage{subfigure}
\usepackage{epstopdf}

\begin{document}
\preprint{UH Preprint: \today}

%Title of paper
\title{Estimating the conditions for polariton condensation in organic thin-film microcavities}

\author{Eric R. Bittner}
\email[]{Bittner@UH.edu}
\homepage[]{http://k2.chem.UH.edu}
%\thanks{}
%\altaffiliation{}
\affiliation{Department of Chemistry, University of Houston, Houston, TX 77204}

%
%\author{Svitlana Zaster}
%%\email[]{Bittner@UH.edu}
%%\homepage[]{http://k2.chem.UH.edu}
%%\thanks{}
%%\altaffiliation{}
%\affiliation{Department of Chemistry, University of Houston, Houston, TX 77204}
%

\author{Carlos Silva}

\affiliation{ Department of Physics and 
Regroupement qu{\'e}b{\'e}cois sur les mat{\'e}riaux de pointe, Universit{\'e} de Montr{\'e}al,\\
C.P. 6128, Succursale centre-ville, \\
Montr{\'e}al (Qu{\'e}bec) H3C 3J7, Canada.}

\date{\today}

\begin{abstract}
We examine the possibility of observing Bose condensation of a confined two-dimensional polariton gas in an organic quantum well.  
We deduce a suitable parameterization of  a model Hamiltonian based upon the cavity geometry, the biexciton binding energy, and 
similar spectroscopic and structural data.   By converting the sum-over-states to a semiclassical integration over $d$-dimensional phase space,
we show that while an ideal 2-D Bose gas will not undergo condensation, an interacting gas with the Bogoliubov dispersion $H(p)\approx s p$ close to $p=0$ will undergo Bose condensation at a given critical density and temperature.  We show that $T_c/\sqrt{\rho_c}$ is 
sensitive to both the cavity geometry and to the biexciton binding energy.  In particular, for strongly bound biexcitons, the non-linear 
interaction term appearing in the Gross-Pitaevskii equation becomes negative and the resulting ground state will be a localized soliton state rather than a delocalized Bose condensate. 
\end{abstract}


\maketitle

Over the past few years, considerable progress has been achieved in creating Bose Einstein condensates (BEC) of weakly interacting 
particles.  The most recent breakthrough is where BEC of photons was realized by pumping an optical cavity filled with a dye and bounded by two reflective mirrors\cite{Klaers:2010fk}. Within the optical cavity, the photons acquire an effective mass as determined by the cut-off frequency of the cavity 
%$m_{eff} = \hbar\omega_{c}/c^{2}$ 
that can be 6-7 orders of magnitude less than mass of an electron.  Depending upon the density, this allows for a BEC transition temperature 
that can approach room temperature.   Polaritons are also ultra-light quasiparticles that are known to condense in systems composed of a 
semiconducting quantum well sandwiched between two reflective mirrors\cite{littlewood:15,wertz:051108,Kasprzak:2006mz,Balili:2007ai}.
In this case, however, the polaritons act as hard-core Bosons and scattering at high density allows for a rapid thermalization of the gas.  Recent
experiments by Kammann {\em et al.} report on the cross-over between photon and polariton BEC within the same microcavity device.\cite{kammann:2011}.

In this paper, we consider the formation of a polariton BEC in an {\em organic} quantum well consisting of a layer of a semiconducting 
polyacene molecular crystal sandwiched between two reflective mirrors.   In contrast to inorganic quantum wells, organic quantum wells can be 
fabricated on the bench-top without the need for ultraclean environments.
Recent experiments by Forrest's group indicate that exciton polaritons in organic quantum wells can exhibit a lasing transition at room temperature\cite{Kena-CohenS.:2010fk,Kena-Cohen:2008tw}, and our recent work indicated that parameteric amplification and the transition to superfluid states  should be possible in these systems even in the presence of
some disorder within the crystal lattice itself \cite{Bittner:2011uq}.
Furthermore, as we shall discuss, the fact that molecular crystals 
may be glassy or have finite-sized microcrystaline domains breaks translation symmetry and may enhance the local density of polaritons thereby resulting in a 
higher BEC transition temperature.    We consider the nature of the polariton/polariton scattering interaction by relating the 
biexciton binding energy to the $S$-wave scattering length for excitons in the cavity. 
Finally, taking molecular-level data as input to our theory, we estimate the transition temperature and critical density 
needed to achieve polariton BEC in a quasi-two dimensional molecular crystal. 

Polaritons are quasiparticles formed by strong coupling between a photon field and excitations within a material.  Within the Frenkel exciton model, we can write a 
model Hamiltonian for this assuming that excitations are local to molecular sites, that excitons can hop to other sites via dipole-dipole coupling of their 
transition moments, and that the exciton/photon interaction in mediated via single quanta exchanges. 
Since polaritons are very lightweight quasi-particles, it is best to work in the long-wave, continuum limit rather than within a molecular-site 
representation more suitable for Frenkel excitons.  However, to do so, we need to determine the effective mass of both the excitons and cavity photons. 
The coherent motion of Frenkel excitons between sites is determined  by coupling between transition dipoles on different molecular sites,
%
%We introduce the relative and center of mass variables ${\bf r}  =({\bf r}_{i}+{\bf r}_{j})/2 $ and ${\bf s} = {\bf r}_{i}-{\bf r}_{j}$ and write 
%$\langle b_{i}^{\dagger}b_{j}\rangle = \langle b_{{\bf r}+{\bf s}/2}^{\dagger} b_{{\bf r}-{\bf s}/2}\rangle$.   
%Thus, one can define the Wigner distribution for the excitons as\cite{Mukamel:1995}
%$$
%W({\bf r},{\bf p}) = \frac{1}{\sqrt{N}}\sum_{{\bf s}}\langle b_{{\bf r}+{\bf s}/2}^{\dagger}b_{{\bf r}-{\bf s}/2}\rangle e^{i{\bf p}\cdot {\bf s}}
%$$
%For the moment, ignoring the interaction with the photon field and collisions between excitons the exciton dynamics are given by 
%$$
%\frac{d}{dt}W({\bf r},{\bf p}) = 2 \sum_{{\bf a}}J({\bf a})\sin({\bf p}\cdot{\bf a})W({\bf r}+{\bf a}/2,{\bf p})
%$$
%This describes the coherent motion of excitons on the molecular lattice.  
$J(r)$.
Since the dipole/dipole interaction is short ranged and centrosymmetric $J(r) = J(-r)$,
 the coherent motion of the exciton in the Wigner representation reduces to 
 $\dot W({\bf r},{\bf p})=({\bf p}/m_{ex})\cdot \vec\nabla_{r}W(r,p)$
  where the 
effective mass of the exciton is given by summing over all sites
$
m_{ex}^{-1} = \sum_{n}J(r_{n})r_{n}^{2}/A,
$
where $A$  is the area of the 2D slab\cite{Mukamel:1995}.
This allows us to readily compute the exciton effective mass given an instantaneous arrangement of the molecules within the 2D quantum well. Using 
the experimental crystallographic arrangement of anthracene, a $S_{o}\to S_{1}$ vertical transition moment of $\mu_{i} =   1.1901 e \text{ bohr}$ and a dielectric constant of $\epsilon = 2.99~$\cite{anthracenedata} we obtain an effective mass of  $m_{ex} = 6.83 m_{e}$ which is close to the value of 10$m_{e}$ typically reported for polyacene molecular crystals.\cite{Agranovich:2001ly} Thus, we write the exciton dispersion as  $$
\hbar\omega_{k}^{ex} = \hbar\omega_{o}^{ex}  + \frac{\hbar^{2}}{2m_{ex}}k^{2}
$$
where $\hbar\omega_{o}^{ex}$ is the exciton transition energy.  For anthracene, we take this to be 3.1117 eV corresponding to the 
experimental vertical absorption energy\cite{Wolf:1975uq}.
Within the cavity, the photon dispersion is given by 
$$
\hbar\omega_{k}^{cav} = \frac{\hbar c}{\eta}\sqrt{k^{2} + \left(\frac{\pi}{L}\right)^{2}},
$$
where $L$ is spacing between the cavity mirrors and $\eta = \sqrt{\epsilon}$ is the refractive index.  If we expand this for small values of $k$, one obtains the parabolic dispersion
\begin{eqnarray}
\hbar\omega_{k}^{cav} = \Delta + \frac{\hbar^{2}}{2m_{eff}}k^{2},\label{eq:cav}
\end{eqnarray}
where $\Delta = 2 \pi \hbar c/L\eta$  is the cut-off energy for the cavity and $m_{eff} = 2 \pi \eta/c L$ is the effective photon mass. 
Taking $L = 250 $nm, $\eta = 2.99$\cite{anthracenedata}  one obtains $\Delta = 2.87eV$ and $m_{eff}/m_{e} = 1.67 \times 10^{-5}$.   
These conditions put the cavity cutoff in resonance with the exciton transition energy, thereby producing the strongest coupling between the 
photon field and the excitons.

\begin{figure}[t]
\includegraphics[width=\columnwidth]{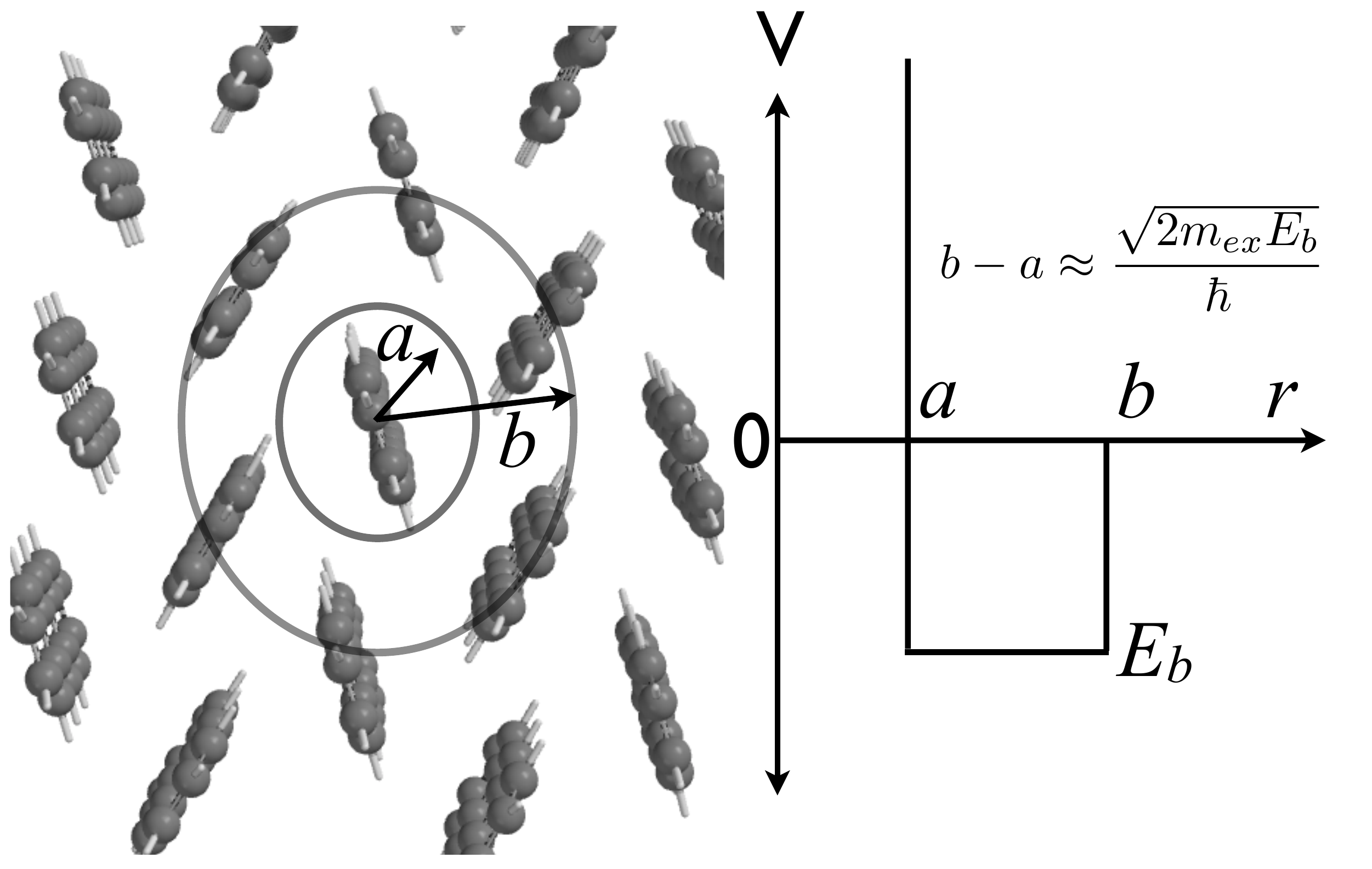}
\caption{Anthracene lattice layer from MD simulation.  The repulsive core radius, $a$, and biexciton binding region are indicated by the concentric 
circles.  The sketch to the right indicates the effective square-well potential for exciton/exciton scattering.}
\label{fig:lattice}
\end{figure}

In the limit of low density,  the combined exciton/cavity system is described by the energy functional
\begin{eqnarray}
{\cal H} = \int d{\bf r} \hat\Psi^{\dagger}({\bf r}).\hat H_{o}.\hat\Psi({\bf r}) + g_{x} \int d{\bf r}(\hat\psi_{x}^\dagger({\bf r}))^{2}(\hat \psi_x({\bf r}))^{2},
\label{hami}
\end{eqnarray}
where $\hat\Psi^{\dagger}({\bf r}) = (\hat\psi^{\dagger}_{ex}({\bf r} ), \hat \psi^{\dagger}_{cav}({\bf r}) )$ is a two-component field containing both exciton and photon 
contributions and
\begin{eqnarray}
\hat H_{o} = \hbar \left(
\begin{array}{cc}
\hat\omega_{x} & \Omega/2 \\
\Omega/2 &  \hat\omega_{cav}  
\end{array} 
\right) + V(r),
\end{eqnarray}
describes the mixing between the exciton and photon fields.   Note, that we have included a trapping potential, $V(r)$ that may 
include any external effect that breaks translational symmetry. 

Polariton/Polariton kinematic interactions were studied in detail by Zoubi and La Rocca \cite{Zoubi:2006kx} using the 
Agranovich-Toshich transformation \cite{Agranovich:1968vn} which converts the Frenkel excitons to Bosons, thereby allowing the
exciton/exciton scattering interaction to be derived.   In short, they show that in the long-wave limit and in the case where the 
LP dispersion is nearly parabolic, the scattering $T$-matrix between polaritons is identical to that scattering from a square well with radius $a$.
With this in mind, we estimate polariton/polariton coupling by assuming that this is due to exciton/exciton scattering.   However, 
here we shall construct an effective square-well potential for exciton/exciton scattering based upon the molecular geometry and available
spectroscopic data. 

The non-linear coupling coefficient, $g_{x} = 4 \pi\hbar \alpha/m_{ex}$ 
is determined by the low-energy S-wave exciton/exciton scattering length, $\alpha$. 
To provide a robust estimation of this term, based upon molecular and spectroscopic properties, 
we assume that stable bi-exciton states exist as a single bound state 
in an effective square well potential with binding energy $\varepsilon_{B}$ and that the range is determined by the nearest neighbor
spacing in the molecular lattice, $b$ as sketched in Fig.~\ref{fig:lattice}.  
Since excitons are Fermions in that we must restrict their population to at most a single exciton per local anthracene site,
 we need to include a hard-core contact radius, $a$, which we take to be the van der Waals radius of anthracene in the molecular crystal plane. 
From quantum scattering theory we arrive at a simple expression for exciton/exciton S-wave scattering length
$$
\alpha = a + (b-a)\left(1-\frac{\tan((b-a)\lambda_{b})}{(b-a)\lambda_{b}}\right),
$$
where $\lambda_{b} = \sqrt{2 m_{ex}\varepsilon_{b}}/\hbar$.    Since $\alpha$ may be either positive or negative depending upon the 
the bi-exciton binding energy, it is possible that different organic media within the cavity would give rise to either bright ($\alpha <0$) 
or dark soliton ($\alpha>0$) ground states. Taking $b = 2a$, the crossover between bright and dark soliton states would occur 
when the de Broglie wavelength of bound biexciton state: $\Lambda_{b}=\hbar/\sqrt{2m_{ex}\varepsilon_{b}} \approx a/\sqrt{3}$.   For systems such as 
anthracene with $\varepsilon_{b}\approx 100 {\rm cm}^{-1}$ and taking $a$ to be 3.5\AA, $\Lambda_{b}/a \approx 2.2$ indicating that
we would expect that the hard-core interaction to dominate, producing  $\alpha>0$ for most polyacene systems.   It is necessary to point out, however, that
the presence of an attractive long-range exciton/exciton interaction does not immediately imply that there will be a bound bi-exciton state.  
It is readily shown that in 1 and 2 dimensions, an arbitrarily shallow square well will always support at least 1 
bound state; however, in 3 dimensions, this is not the case. 

Polaritons are formed by transforming $H_{o}$ into a diagonal representation, yielding upper and lower branches with dispersion
$$
\omega_{L,U} (k)= \frac{\omega_{k}^{cav}+\omega_{k}^{ex}}{2} \pm \frac{1}{2}\sqrt{\Omega^{2} +(\omega_{k}^{cav}-\omega_{k}^{ex})^{2}},
$$
where $\Omega$ is the Rabi frequency and $\theta_{k} ={\Omega}/(\omega_{k}^{cav}-\omega_{k}^{ex})$ is the mixing angle. 
If we ignore mixing between the upper and lower branches (due to exciton/exciton scattering),  the ground state of the system is a 
stationary solution of the Gross-Pitaevskii (GP) equation  \cite{Pitaevskii:1961zr,Gross:1961ys},
\begin{eqnarray}
\left(
\hbar\hat\omega_{LP} + V(r) +g |\phi_{c}(r)|^{2}\right)\phi_{c}(r) = \mu \phi_{c}(r),
\end{eqnarray}
where $\mu$ is the chemical potential.  In this expression, the $\phi_{c}$ is normalized to unity
since we have included the number of excitations in system into the non-linearity, $g = g_{x} X_{o}^{2} N_{cav},$
where $X_{o}$ is the Hopfield coefficient for the exciton portion of the LP wave function evaluated at $k=0$ and $N_{cav}$ is the number of photons in the cavity.

%Fig.~\ref{fig1} shows the  dispersion curves for a $L = 250$nm 
%anthracene cavity.  Close to the bottom of the LP curve, the dispersion is nearly parabolic and we can determine the 
%effective lower polariton mass by expanding
%\begin{eqnarray}
%\omega_{L}(k) \approx \omega_{L}(0)  + \frac{\hbar^{2}}{2m_{L}}k^{2} + \cdots \label{eq:lp}
%\end{eqnarray}
%For the case at hand, we obtain an effective lower polariton mass of $m_{L} = 0.00019 m_{e}$.
%\footnote{An interactive calculator for computing cavity parameters is available as a Wolfram Demonstration.}
%\begin{figure}
%\includegraphics[width=\columnwidth]{Fig1}
%\caption{Polariton dispersion curves based upon anthracene lattice model.}
%\label{fig1}
%\end{figure}

For an ensemble of Bosons, the average number of particles in the gas is given by 
\begin{eqnarray}
N %&=& \sum_{k}\frac{\lambda e^{-\beta \epsilon_k}}{(1-\lambda e^{-\beta\epsilon_k})} \\
&=&\frac{\lambda}{1-\lambda} +  \sum_{k\ne0}\frac{\lambda e^{-\beta \epsilon_k}}{(1-\lambda e^{-\beta\epsilon_k})} \label{eq:pop} \\
&=& n_{o} + N_{ex}, \nonumber
\end{eqnarray}
where  $\lambda = \exp(\mu\beta) $ and we have pulled out the term representing the population of the ground state. 
The sum in $N_{ex}$ is over all excitation of the system.  
An ideal gas  of bosons will not form a condensate
unless the dimensionality of the system, $d > 2$.  The reason for this is that density of states scales as 
$\omega(\epsilon)d\epsilon \propto \epsilon^{d/2-1}d\epsilon$ and consequently the sum over the 
excited states diverges and BEC is forbidden by translational symmetry in 1 and 2 dimensions for a non-interacting system.

In the experiments by Snoke and coworkers
\cite{Balili:2007ai,DavidSnoke11152002}, translational symmetry is broken by putting a local strain on the cavity.  
It has also been argued that the finite size of the  pumping laser is sufficient to break the 2D translational symmetry.
Likewise, in the photon condensation reported by Klaers {\rm et al.}, condensation occurred within a finite 2D cavity
with parabolic mirrors\cite{Klaers:2010fk}. 
 For the case of a molecular crystal, translational symmetry can be broken by the presence of 
 microcystaline domains within the sample itself.  Within a given domain, the electronic transition moments are 
 aligned along the crystallographic axes. The grain boundaries impose a hard-wall trapping potential.  This should allow for a much higher density of polaritons since excitons created within a given domain will be confined to that domain rather than being free to diffuse through out the sample. 
 However, as we show next, grain boundaries alone are insufficient to break 2D symmetry and one needs to consider the effect of the 
 interactions. 

In order to estimate the transition temperature for a polariton gas in an organic microcavity, let us assume that the
number of polaritons is high enough that we can ignore the kinetic energy term in the GP equation and that $V(r)$ is a 
circular well potential of radius $R_o$ corresponding to the typical size of an crystalline microdomain.   Within the 
well, we can write $\phi_c = (1/(2\pi R_o^2))^{1/2}$  and thus, $\mu = g /(2\pi R_o^2)$ (recall, that $g$ already includes the 
number of quasi-particles in the system).    Excitations of the condensate are
then determined by writing $\hat\psi =\phi_{c} + \delta\psi$ and reintroduce this into the GP equation. 
Assuming the energy dispersion for the non-interacting system is quadratic in $k$
the energy dispersion for the excitation is given by the Bogoliubov dispersion
$$
\epsilon_k = \left(\frac{\hbar^2}{2m_L}k^2\left(\frac{\hbar^2}{2m_L}k^2 + 2 g\right)\right)^{1/2}.
$$
For $g$ sufficiently large, we can  approximate the Bogoliubov spectrum as 
$\epsilon_k = s |p|$ where $s$ is the speed of sound $s = (g/m_{L})^{1/2}$ and $p =\hbar k$ is the momentum. 
In this  regime where the dispersion is linear in $p$, the condensate is behaving as  a superfluid according to the Landau criterion. 

\begin{figure}
\includegraphics[width=\columnwidth]{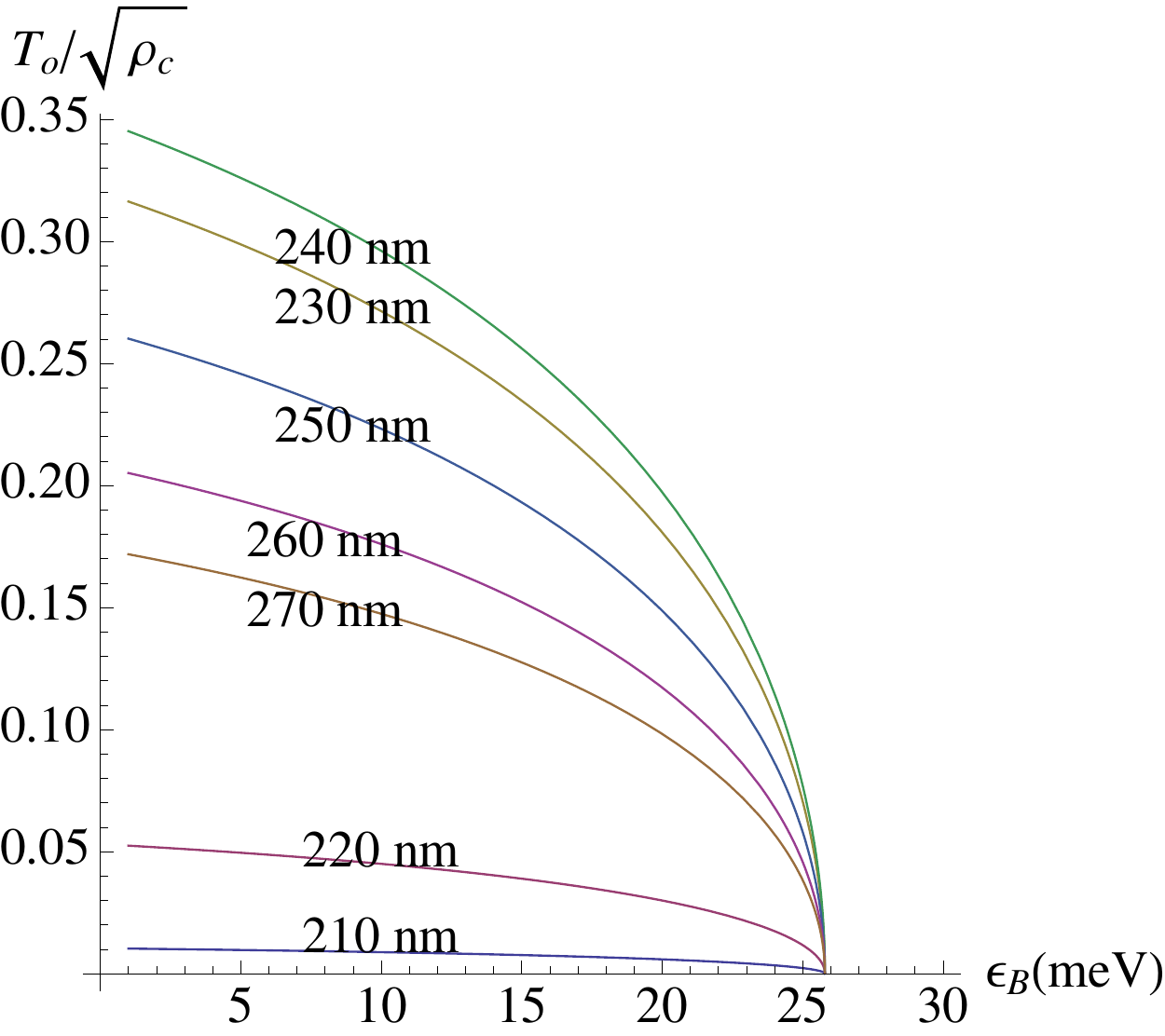}
\caption{Scaled BEC transition temperature versus biexciton binding energy for various cavity sizes  }
\label{fig3}
\end{figure}

To  evaluate $N_{ex}$, let us write the classical Hamiltonian for a particle in a spherical  potential well with radius $R_o$  using 
$H(p,r) =s p^a $ and  invoke the  semi-classical approximation.  
For $a = 1$ we have the linear dispersion for the Bogoliubov excitation and for $a = 2$ we have the ordinary quadratic dispersion. 
Thus,  we can convert the sum over states to a integral over phase space.  
\begin{eqnarray}
\sum_{k\ne 0} \to \frac{d^2c_d^2}{h^d}
\int_{0}^{\infty}
p^{d-1} dp
\int_{0}^{R_o}r^{d-1} dr,
\end{eqnarray}
with $c_d =\pi^{d/2}/\Gamma(d/2+1)$.    The resulting integral converges for all dimensions and 
positive values of $U$ and $d$.    In particular, 
 for a 2D cavity and $a =1$ with radius $R_o$,
 \begin{eqnarray}
\frac{N_{ex}}{A}=
\frac{2 c_2 \text{Li}_2(\lambda )}{\beta ^2 h^2 s^2}, \label{eq:nex} 
\end{eqnarray}
where $A$ is the surface area of the microdomain and $\text{Li}_n(x)$ is the polylog function. 
Eq.~\ref{eq:nex}  allows us to predict the transition temperature for  the formation of a polariton BEC using input strictly from molecular or spectroscopic considerations.  It follows that, 
\begin{eqnarray}
\frac{n_o}{N} = 1-\left(\frac{T}{T_o}\right)^2
\end{eqnarray}
for $T < T_o$ and 
\begin{eqnarray}
k_BT_o &=& \left(\frac{3}{\pi^3}\right)^2 \rho^{1/2} hs.
\end{eqnarray}
It is important to note that the cavity properties and number of cavity photons enters into this relation in two ways.  First in the density, $\rho \approx
 X_o^2 \rho_{cav}$ and in the sound velocity, $s$. 
Fig.~\ref{fig3} shows the variation  in the transition temperature (scaled by $\sqrt{\rho_{cav}}$) for various values of the biexciton binding energy
and the cavity size.   The variation with cavity size  reflects the exciton fraction of the polariton wavefunction.  Close to resonance, the 
Hopfield coefficient, $X_o^2$, is at its maximum value. 
It is interesting to note that as the biexciton binding increases, the transition temperature generally decreases.  This is
because the attractive part of the exciton/exciton interaction decreases the effective exciton scattering length.   However, for strongly bound
bound bi-excitons, the scattering length can become negative resulting in an transition from a BEC state to a bright-soliton state.  
In this case, $T_o$ becomes imaginary and the transition temperature can not be defined.   For the model cavity at hand, this 
occurs when $\varepsilon_B = 26.6$meV.   This is a large biexciton binding energy for organic molecular crystal systems.

In conclusion, we predict that polariton BEC should be readily observable in an organic microcavity system using polyacene thin films.  The arguments we present are based entirely up either molecular or spectroscopic parameters.  Our analysis is based upon 
the notion that the non-linear interaction in the GP equation can be deduced from the biexciton scattering which we treat 
in terms of a repulsive inner core surrounded an attractive square-well interaction.   For strongly bound biexcitons, the 
exciton/exciton scattering length can become negative and the resulting polariton ground state will be a bright-soliton rather than a 
Bose Einstein condensate.   The results presented here are not limited to organic molecular crystals and apply equally to $J$-aggregate systems, such as those recently studied by the Bulovi{\`c} group at MIT~\cite{PhysRevB.78.193305}.

\begin{acknowledgments}
The work at the University of Houston was funded in part by the National Science Foundation (CHE-1011894) and the Robert A. Welch Foundation (E-1334).
CS acknowledges support from the Canada Research Chair in Organic Semiconductor Materials.
\end{acknowledgments}

%\bibliography{/Users/ebittner/Documents/References/References}

\end{document}